\documentclass[twocolumn,prl,showpacs]{revtex4}
\usepackage{graphicx}
\usepackage{dcolumn}
\usepackage{bm}
\usepackage{amsmath}

\begin{document}

\preprint{}
\title{Dynamics of magnetization on the topological surface}
\author{Takehito Yokoyama$^{1}$, Jiadong Zang $^{2,3}$, and Naoto Nagaosa$%
^{2,4}$}
\affiliation{$^1$Department of Physics, Tokyo Institute of Technology, Tokyo 152-8551,
Japan \\
$^2$Department of Applied Physics, University of Tokyo, Tokyo 113-8656,
Japan \\
$^3$Department of Physics, Fudan University, Shanghai 200433, China\\
$^4$Cross Correlated Materials Research Group (CMRG), ASI, RIKEN, WAKO
351-0198, Japan }
\date{\today}

\begin{abstract}
We investigate theoretically the dynamics of magnetization coupled to
the surface Dirac fermions of a three dimensional topological insulator, by
deriving the Landau-Lifshitz-Gilbert (LLG) equation in the presence of 
charge current. Both the inverse spin-Galvanic effect and the Gilbert damping
coefficient $\alpha$ are related to the two-dimensional diagonal conductivity
$\sigma_{xx}$ of the Dirac fermion, while the Berry phase of the
ferromagnetic moment to the Hall conductivity $\sigma_{xy}$. 
The spin transfer torque and the so-called $\beta$-terms are shown to be negligibly small. 
Anomalous behaviors in various
phenomena including the ferromagnetic resonance are predicted in terms of
this LLG equation.
\end{abstract}

\maketitle


Topological insulator (TI) provides a new state of matter topologically
distinct from the conventional band insulator~\cite{Hasan}. In particular, the
edge channels or the surface states are described by Dirac fermions and
protected by the band gap in the bulk states, and backward scattering is
forbidden by the time-reversal symmetry. From the viewpoint of the
spintronics, it offers a unique opportunity to pursue novel functions since
the relativistic spin-orbit interaction plays an essential role there.
Actually, several proposals have been made such as the quantized
magneto-electric effect~\cite{Qi}, giant spin rotation~\cite{Yokoyama1},
magneto-transport phenomena~\cite{Yokoyama2}, and superconducting proximity
effect including Majorana fermions \cite{Fu,Tanaka,Linder}.

Also, a recent study focuses on the inverse spin-Galvanic effect in a TI/ferromagnet
interface, predicting the current-induced magnetization reversal due to the
Hall current on the TI~\cite{Garate}. In Ref.~\cite{Garate}, the Fermi
energy is assumed to be in the gap of the Dirac dispersion opened by the
exchange coupling. In this case, the quantized Hall liquid is realized, and
there occurs no dissipation coming from the surface Dirac fermions.

However, in realistic systems, it is rather difficult to tune the Fermi
energy in the gap since the proximity-induced exchange field is expected to be around 5-50meV. Therefore, it is
important to consider the generic case where the Fermi energy is at the
finite density of states of Dirac fermions, where the diagonal conductivity
is much larger than the transverse one, and the damping of the magnetization
becomes appreciable. Related systems are semiconductors and metals with
Rashba spin-orbit interaction, where the spin-Galvanic
effect and current induced magnetization reversal have been predicted~\cite%
{Manchon} and experimentally observed~\cite{Chernyshov,Miron}. Compared with
these systems where the Rashba coupling constant is a key parameter, the
spin and momentum in TI is tightly related to each other corresponding to
the strong coupling limit of spin-orbit interaction, and hence the gigantic
spin-Galvanic effect is expected.

In this letter, we study the dynamics of the magnetization coupled to the
surface Dirac fermion of TI. Landau-Lifshitz-Gilbert (LLG) equation in the presence of charge current is derived microscopically, and (i) inverse spin-Galvanic  effect, (ii) Gilbert damping coefficient $\alpha$, (iii) the so-called $\beta$-terms, and (iv) the correction to the Berry phase, are derived
in a unified fashion. It is found that these are expressed by relatively
small number of parameters, i.e., the velocity $v_F$, Fermi wave number $k_F$%
, exchange coupling $M$, and the transport lifetime $\tau$ of the Dirac
fermions. It is also clarified that the terms related to the spatial gradient are negligibly small when the surface state is a good metal. With this LLG
equation, we propose a ferromagnetic resonance (FMR) experiment, where 
modifications of the resonance frequency and Gilbert damping 
are predicted. 
Combined with the transport measurement of the Hall conductivity, FMR provide several tests of our theory.

\begin{figure}[tbp]
\begin{center}
\scalebox{0.8}{
\includegraphics[width=9.5cm,clip]{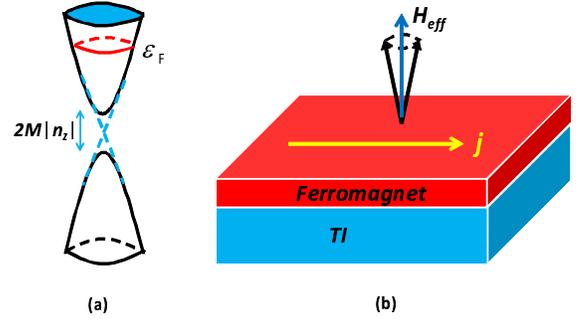}
}
\end{center}
\caption{(Color online) (a) Illustration of the Dirac dispersion on top of TI. The Fermi level $\protect\varepsilon_F$ is far above
the surface gap opened by magnetization in the ferromagnetic
layer. (b) Sketch of FMR experiment in the soft magnetic layer. The
substrate in the figure is TI, which is capped by
a layer of soft ferromagnet. The magnetization precesses around the
external magnetic field ${\bm H}_{\mathrm{eff}}$.}
\label{fig1}
\end{figure}

\textit{Derivation of LLG equation.} --- 
By attaching a ferromagnet on the TI as shown in Fig. \ref{fig1}, we can consider a topological surface state where conducting electrons interact with localized
spins, ${\bm S}$, through the exchange field
\begin{equation}
H_{\mathrm{ex}}=-M\int d{\bm r}\,{\bm n}({\bm r})\!\cdot \!\hat{\bm\sigma }({%
\bm r}).
\end{equation}%
Here, we set ${\bm S}=S{\bm n}$ with a unit vector ${\bm n}$ pointing in the
direction of spin, $\hat{\bm\sigma }({\bm r})=c^{\dagger }({\bm r}){\bm%
\sigma }c({\bm r})$ represents (twice) the electron spin density, with $%
c^{\dagger }=(c_{\uparrow }^{\dagger },c_{\downarrow }^{\dagger })$ being
electron creation operators, ${\bm\sigma }$ the Pauli spin-matrix vector,
and $M$ being the exchange coupling energy. The total Hamiltonian of the
system is given by $H_{\mathrm{tot}}=H_{S}+H_{\mathrm{el}}+H_{\mathrm{ex}}$,
where $H_{S}$ and $H_{\mathrm{el}}$ are those for localized spins and
conducting electrons, respectively.

The dynamics of magnetization can be described by the
LLG equation
\begin{eqnarray}
\dot {\bm n} &=& \gamma_0 {\bm H}_{\mathrm{eff}} \times {\bm n} + \alpha_0
\dot {\bm n} \times {\bm n} + {\bm t}_{\mathrm{el}}^{\prime},  \label{eq:LLG}
\end{eqnarray}
where $\gamma_0{\bm H}_{\mathrm{eff}}$ and $\alpha_0$ are an effective field
and a Gilbert damping constant, respectively, both coming from $H_S$.
Effects of conducting electrons are contained in the spin torque
\begin{eqnarray}
{\bm t}_{\mathrm{el}} ({\bm r}) &\equiv& s_0 \ {\bm t}_{\mathrm{el}%
}^{\prime}({\bm r}) \ = \ M {\bm n} ({\bm r}) \times \langle \hat {\bm \sigma%
} ({\bm r}) \rangle_{\mathrm{ne}},  \label{eq:torque0}
\end{eqnarray}
which arises from $H_{\mathrm{ex}}$. Here, $s_0 \equiv S/a^2$ is the
localized spin per area $a^2$. In the following, we thus calculate
spin polarization of conducting electrons perpendicular to ${\bm
n}$, $\langle \hat{\bm \sigma}_\perp ({\bm r})
\rangle_{\mathrm{ne}}$, in such nonequilibrium states with current
flow and spatially varying magnetization to derive the $\beta$-term, or with
time-dependent magnetization for Gilbert damping. Here and hereafter, $%
\langle \cdots \rangle_{\mathrm{ne}}$ represents statistical average in
such nonequilibrium states.

Following Refs. \cite{Kohno,TBB,TFH} we consider a small transverse
fluctuation, ${\bm u} = (u^x,u^y,0)$, $|{\bm u}| \ll 1$, around a uniformly
magnetized state, ${\bm n} = \hat z$, such that ${\bm n} = \hat z + {\bm u}$%
. In the \lq unperturbed' state, ${\bm n} = \hat z$, the electrons are
described by the Hamiltonian
\begin{eqnarray}
\mathcal{H}_0 &=& \sum_{{\bm k}} v_F \left( {k_y \sigma ^x - k_x \sigma ^y }
\right) - M\sigma ^z - \varepsilon _F + V_{\mathrm{imp}}  \label{eq:H0}
\end{eqnarray}
where $V_{\mathrm{imp}}$ is the impurity potential given by $V_{\mathrm{imp}%
} = u \sum_i \delta ({\bm r} - {\bm R}_i) $ 
in the first-quantization form. We take a quenched average for the impurity
positions ${\bm R}_i$. The electron damping rate is then given by $\gamma =
1/(2\tau)= \pi n_{\mathrm{i}}u^2 \nu_F$ 
in the first Born approximation. Here, $n_{\mathrm{i}}$ is the concentration
of impurities, and $\nu _F = \varepsilon _F / (2 \pi v_F^2) $ is the density
of states at $\varepsilon_{\mathrm{F}}$. We assume that $\gamma \ll v_{%
\mathrm{F}} k_{\mathrm{F}} = \sqrt{\varepsilon _F^2 - M^2}, M$, and
calculate spin transfer torque in the lowest non-trivial order.

In the presence of ${\bm u}({\bm r},t) = {\bm u}({\bm q},\omega) \, \mathrm{e%
}^{i({\bm q}\cdot{\bm r}-\omega t)}$, the conducting electrons feel a
perturbation (note that $H_{\mathrm{el}} + H_{\mathrm{ex}} = \mathcal{H}_0 +
\mathcal{H}_1$)
\begin{eqnarray}
\mathcal{H}_1 &=& - M \sum_{{\bm k} \sigma} c^\dagger_{{\bm k}+{\bm q}} {\bm %
\sigma} c^{\phantom{\dagger}}_{{\bm k}} \!\cdot\! {\bm u} ({\bm q},\omega)
\, \mathrm{e}^{-i\omega t} ,  \label{eq:H1}
\end{eqnarray}
and acquires a transverse component
\begin{eqnarray}
\langle \hat \sigma_\perp^{\prime \alpha} ({\bm q}, \omega ) \rangle_{%
\mathrm{ne}} &=& M \chi_\perp^{\alpha \beta} ({\bm q}, \omega + i0) \,
u^\beta ({\bm q},\omega)
\end{eqnarray}
in the first order in ${\bm u}$ in the momentum and frequency
representation. Here, $\chi_\perp^{\alpha \beta}$ is the transverse spin
susceptibility in a uniformly magnetized state with $\alpha, \beta = x,y$,
and summing over $\beta$ is implied.

Now, we study the $\omega$-linear terms in the uniform (${\bm q}={\bm 0}$)
part of the transverse spin susceptibility, $\chi_\perp^{\alpha \beta} ({\bm %
q}={\bm 0}, \omega + i0)$. We make the following transformation of the
operator:
\begin{eqnarray}
c = U\tilde c = \frac{1}{{\sqrt {2\varepsilon (\varepsilon + M)} }}\left( {%
\begin{array}{*{20}c} { v_F (k_y + ik_x )} \\ {\varepsilon + M} \\
\end{array}} \right)\tilde c
\end{eqnarray}
with $\varepsilon = \sqrt {(v_F k)^2 + M^2 }$. Note $U^\dag U = 1, U^\dag
\sigma ^x U = v_F k_y /\varepsilon,$ and $U^\dag \sigma ^y U = -v_F k_x
/\varepsilon$. This transformation maps two component operator $c$ into one
component operator on the upper Dirac cone $\tilde c$. With this new
operator, we calculate the transverse spin susceptibility in Matsubara form
\begin{eqnarray}
\chi _ \bot ^{\alpha \beta } (\mathbf{0},i\omega _\lambda ) = \int_0^\beta {%
d\tau e^{i\omega _\lambda \tau } \left\langle {T_\tau \sigma ^\alpha (%
\mathbf{0},\tau )\sigma ^\beta ( \mathbf{0},0)} \right\rangle }  \notag \\
= - T\sum\limits_{\mathbf{k},n} {U^\dag \sigma ^\alpha U\tilde G(\mathbf{k}%
,i\varepsilon _n + i\omega _\lambda )} U^\dag \sigma ^\beta U\tilde G(%
\mathbf{k},i\varepsilon _n )
\end{eqnarray}
with $\tilde G(\mathbf{k},i\varepsilon _n ) = \left( {i\varepsilon _n -
\varepsilon + \varepsilon _F + i\gamma {\mathop{\rm sgn}} (\varepsilon _n )}
\right)^{ - 1}$. 
By symmetry consideration of the integrand in ${\bm k}$-integral, we find $\chi
_ \bot ^{\alpha \beta } (0,i\omega _\lambda ) \propto \delta _{\alpha \beta
} $. After some calculations, we obtain the torque
stemming from the time evolution:
\begin{eqnarray}
\mathbf{t}_{el}^\alpha = M^2 \frac{{i\omega }}{{2\pi }}\frac{1}{{2v_F^2 }}%
\left( {\frac{{v_F k_F^{} }}{{\varepsilon _F }}} \right)^2 \varepsilon _F
\tau \mathbf{n} \times \mathbf{u} \\
= \frac{1}{2}\left( {\frac{{Mv_F k_F^{} }}{{\varepsilon _F }}} \right)^2 \nu
_F \tau \mathbf{\dot n} \times \mathbf{n} .
\end{eqnarray}
This result fits the conventional Gilbert damping with
\begin{equation}  \label{eq:alpha}
\alpha =\frac{1}{2}\left( {\frac{{Mv_{F}k_{F}^{{}}}}{{\varepsilon _{F}}}}%
\right) ^{2}\nu _{F}\tau \frac{{a^{2}}}{\hbar S}.
\end{equation}%
%
%


We next examine the case of finite current by applying a d.c. electric field
${\bm E}$, and calculate a linear response of $\sigma_\perp^\alpha$ to ${\bm %
E}$, i.e., $<\sigma_\perp^\alpha ( {\bm q})>_{\rm ne} = K_i^\alpha
({\bm q}) E_i$. First, it is clear that $K_i^{\alpha} ({\bm q}={\bm
0}) = -\varepsilon_{i \alpha} \sigma_{xx}/(e v_F)$ where
$\varepsilon_{i \alpha}$ and $\sigma_{xx}$ are the anti-symmetric tensor and diagonal conductivity, respectively, because
electron's spin is "attached" to its momentum. This represents the inverse
spin-Galvanic effect, i.e., charge current induces magnetic
moment. Since we assume that Fermi level is
far away from the surface gap, $\sigma_{xx}\gg\sigma_{xy}$ where $\sigma_{xy}$ is the Hall conductivity. 
The dominant term in $\chi$ is thus $\chi_{xy}\propto\sigma_{xx}$.
This is quite different from the case studied in Ref. \cite{Garate},
where Fermi level lies inside the surface gap and therefore
$\sigma_{xx}$ is vanishing. Hence, the only contribution to the inverse
spin-Galvanic effect is $\chi_{xx}\propto\sigma_{xy}$, which is much
smaller than the effect proposed in this letter. Compared with the inverse
spin-Galvanic effect in Rashba
system~\cite{Manchon,Chernyshov,Miron}, this effect is much stronger
since the small Rashba coupling constant, i.e., the small factor
$\alpha_R k_F/E_F$ in Eq. (16) of Ref.~\cite{Manchon}, does not
appear in the present case. Taking into account the realistic
numbers with $\alpha=10^{-11}eVm$ and $v_F=3\times10^5m/s$, one finds that
the inverse spin-Galvanic effect in the present system is $\sim$ 50 times larger
than that in Rashba systems.

The next leading order terms of the expansion in $u^\beta$ and $q_j$ can be 
obtained by considering the four-point vertices \cite{Kohno} as
\begin{widetext}
\begin{eqnarray}
 \langle {\hat\sigma_\perp ^\alpha  ({\bm q}) } \rangle_{\rm ne}  =  - eM\frac{\pi }{4}\frac{{5i}}{{8\pi \varepsilon _F^2 }}\varepsilon _{ik} \varepsilon _{jl} \left[ {\delta _{\alpha \beta } \delta _{kl}  + \delta _{\alpha k} \delta _{\beta l}  + \delta _{\alpha l} \delta _{\beta k} } \right]q_j u^\beta  E_i  \\
  =  - eM\frac{{5i}}{{32\varepsilon _F^2 }}\left[ {{\bf{q}} \cdot {\bf{E}}u^\alpha   - {\bf{q}} \cdot ({\bf{u}} \times {\hat{z}})({\bf{E}} \times {\hat{z}})_\alpha   + {\bf{u}} \cdot ({\bf{E}} \times {\hat{z}})({\bf{q}} \times {\hat{z}})_\alpha  } \right] .
\end{eqnarray}
Therefore, the spin torque steming from the spatial gradient has the form:
\begin{eqnarray}
{\bm t}_{\rm el}^{\beta} =  - \beta \frac{1}{{2e}}\left[ {{\bf{n}} \times ({\bf{j}} \cdot {\bm \nabla} ){\bf{n}} - ({\bf{j}} - ({\bf{j}} \cdot {\bf{n}}){\hat{z}}){\bm \nabla}  \cdot ({\bf{n}} \times {\hat{z}}) + ({\bm \nabla}  - ({\bf{n}} \cdot {\bm \nabla} ){\hat{z}}){\bf{n}} \cdot ({\bf{j}} \times {\hat{z}})} \right]
\label{eq:torque}
\end{eqnarray}
\end{widetext}
where ${\bf{j}} = \sigma_C {\bf{E}}$ with charge current ${\bf{j}}$ and conductivity
$\sigma _C  = \frac{{e^2 }}{{4\pi }}\left( {\frac{{v_F k_F^{} }}{{\varepsilon _F }}}
\right)^2 \varepsilon _F \tau $.
and
\begin{equation}  \label{eq:beta}
\beta =\frac{{5\pi }}{{4\varepsilon _{F}\tau }}\left( {\frac{M}{{%
v_{F}k_{F}^{{}}}}}\right) ^{2}.
\end{equation}%
%
%
%
From Eq.(\ref{eq:torque}), one can find the followings: (i) The spin
transfer torque of the form $(\mathbf{j} \cdot {\bm \nabla}
)\mathbf{n}$ is missing since we consider the upper Dirac cone only.
(ii) The $\beta$-term
has a form essentially different from that in the conventioal one.\cite%
{Kohno,Wong,Zhang} In contrast to the conventional ferromagnet,\cite{Kohno}
this constant comes from the nonmagnetic impurity. Considering $v_{F}k_{F}^{{}}\cong \varepsilon _{F}$, we get $\alpha/\beta \sim (\varepsilon_F
\tau)^{2}$ from Eqs. (\ref{eq:alpha}) and (\ref{eq:beta}). Therefore, the $%
\beta$-terms are negligible for a good surface metal, i.e., $\varepsilon _F \tau \gg 1$.

Up to now, we consider only one branch of the band where the Fermi energy is
sitting. When we consider the 2-band structure, i.e., the 2$\times$2 matrix Hamiltonian  $\mathcal{H}=v_F[(k_y+\frac{Mn_x}{v_F})\sigma^x-(k_x-\frac{%
Mn_y}{v_F})\sigma^y]$, we have the correction to the Berry phase term.
In analogy with the minimal coupling of electromagnetic field, $\mathbf{A}=-%
\frac{M}{ev_F}(-n_y,n_x)$ plays the same role as the $U(1)$ gauge.
By integrating the fermions out, one can get a Chern-Simons term in
terms of the magnetization $L_{CS} =
\sigma_{xy}\epsilon^{\mu\nu\rho}A_\mu\partial_\nu A_\rho $ where $%
\mu,\nu,\rho=t,x,y$. When the gradient of magnetization
vanishes, it can be rewritten as
\begin{equation}
L_{CS}=\sigma_{xy}(\frac{M}{ev_F})^2(n_x\dot{n}_y-n_y\dot{n}_x).
\end{equation}
This additional term can be interpreted as an additional Berry phase for
the magnetization. In fact, as $n_z$ remains constant in the present case,
we have $[n_x,n_y]=in_z$. Therefore, $n_x$ and $n_y$ become conjugate
variables up to a factor, which naturally leads to a Berry phase: $n_x\dot{n%
}_y-n_y\dot{n}_x $. This term is exactly equivalent to the Chern-Simons term.

Including all the terms derived above, we finally arrive at a modified LLG
equation: 
\begin{widetext} 
\begin{eqnarray}
\dot{\bm n} -2\sigma_{xy}(\frac{M}{ev_{F}})^{2}\dot{{\bm {\bm n}}}/(s_0N) = \gamma _{0}{\bm H}_{\mathrm{eff}}\times {\bm n}   + \left( {\frac{M}{{ev_F s_0 N}}} \right)\left( { - {\bf{j}} + ({\bf{n}} \cdot {\bf{j}})\hat z} \right)  
 + (\alpha_{0}+\alpha/N)\dot{\bm n}\times {\bm n}
 +{\bm t}_{\mathrm{el}}^{\beta }/(s_{0} N)
\label{eq:LLG2}
\end{eqnarray}%
\end{widetext}
where $N$ is the number of ferromagnetic layers. Note that $\alpha$-, $\beta$- and Berry phase terms originate from the interplay between Dirac
fermions and local magnetization which persists over a few layers of the ferromagnet. 
Therefore, the overall coefficients are divided by the number of ferromagnetic layers $N$.

\textit{Ferromagnetic resonance.} ---Observing the small value of
$\beta $, the spatial gradient of magnetization can be neglected for the
time being. Only one uniform domain in the absence of current is
taken into account for simplicity. Without loss of generality,
assume that an external magnetic field is applied along $z$
direction, and consider the ferromagnet precession around that field. $\dot{n%
}_{z}=0$ is kept in the first order approximation, namely $n_{z}$ is a
constant in the time evolution. By inserting the ansatz $%
n_{x(y)}(t)=n_{x(y)}e^{-i\omega t}$ into the modified LLG equation, one obtains

\begin{equation}
\Re \omega =\frac{\xi }{\xi ^{2}+\eta ^{2}}\omega _{0},\quad \Im \omega =-%
\frac{\eta }{\xi ^{2}+\eta ^{2}}\omega _{0}  \label{eq:FMR}
\end{equation}%
where $\eta =(\alpha _{0}+\alpha /N)$,  $\omega
_{0}=\gamma _{0}H_{eff}$ and $\xi =1- 2\sigma _{xy}(\frac{M}{ev_{F}%
})^{2}/(s_0N)$. Expanding up to the first order in $\sigma _{xy}$ and $%
\eta $, one gets $\Re \omega =\omega _{0}+2\sigma _{xy}(\frac{M}{%
ev_{F}})^{2}\omega _{0}/( s_0 N)$ and $\Im \omega =\eta \omega
_{0}$. Therefore, the precession frequency acquires a shift
proportional to $\sigma _{xy}$ in the presence of interplay between
Dirac fermions and the ferromagnetic layer. The relative shift of
$\Re\omega$ is $2\sigma _{xy}(\frac{M}{%
ev_{F}})^{2}\omega _{0}/( s_0 N)=\frac{1}{\pi
SN}\frac{M}{\varepsilon_F}(\frac{Ma}{v_F})^2 \sim \frac{1}{N} (\frac{M}{\varepsilon_F})^3$\cite{Zang}. By tuning the
Fermi level, this shift can be accessible
experimentally. 

Meanwhile, the Gilbert damping constant $\alpha $ can be measured
directly without referring to the theoretical expression in Eq. (\ref%
{eq:alpha}). 
One can investigate the ferromagnetic layer thickness dependence of
FMR line-width. While increasing the thickness $N$ of ferromagnet,
the Gilbert damping constant stemming from the Dirac fermions decreases inversely proportional to the thickness. 
Taking into account the realistic estimation with $%
\varepsilon _{F}\tau \sim 100$ and $M/\varepsilon_F \sim 0.3$, one has $\alpha/s_0 \sim 1$, while $%
\alpha _{0} \sim 0.001$ usually. Therefore, even for a hundred of layers of ferromagnet, the contribution from the proximity effect
is still significant compared to the one coming from the ferromagnet
itself. Observing that the imaginary part of resonance frequency
in Eq. (\ref{eq:FMR}) is proportional to $\eta $, one may plot the
relation between the FMR peak broadening, namely $\Im \omega $, and
$1/N$. 
The broadening is a linear function of $1/N$, and approaches the value of the ferromagnet at large thickness limit. We can find the value of $\alpha$ from the slope of the plot.

On the other hand, the real part of FMR frequency provides rich physics as
well. Since in the presence of additional Berry phase, the frequency shift
is proportional to the Hall conductivity on the surface of TI, it leads to a new method to measure the Hall conductivity without
four-terminal probe. 
In an ideal case when the Fermi level lies inside the surface gap, this quantity is quantized as $\sigma_{xy}^0=\frac{e^2}{2h}$. However, in realistic case,
Fermi level is away from the surface gap, and therefore the Hall
conductivity is reduced to $\sigma_{xy}=\frac{e^2}{2h}\frac{Mn_z}{%
\varepsilon_F}$\cite{Zang}. As a result, the shift of resonance frequency is
proportional to $n^2_z\propto \cos^2\theta$, and the FMR isotropy is broken.
Here, $\theta$ is the angle between effective magnetic field and the normal to
the surface of TI. One can perform an angle resolved FMR
measurement. The signal proportional to $\cos^2\theta$ comes from additional
Berry phase.


Since parameters $\alpha $ and $\beta $ depend on $M$ and $\tau $,
it is quite important to measure these quantities directly.
Molecular-beam epitaxy method can be applied to grow TI coated by a thin layer of soft ferromagnet. As is required in the above calculation, Fermi level of TI should lie inside the bulk band gap. Also, the soft ferromagnet should be an insulator or a metal with proper work function. One may employ angular resolved photoemission spectroscopy (ARPES) or scanning tunneling microscope techniques
to measure the surface gap $\Delta $ opened by the ferromagnet, which is
given by $\Delta =Mn_{z}$. As the easy axis $n_{z}$ can be found experimentally, $M$ can be fixed as well. On the other hand, the lifetime $%
\tau $ is indirectly determined by measuring the diagonal conductivity $%
\sigma _{xx}$ via $\sigma_{xx} =\frac{{e^2 }}{{%
4\pi }}\left( {\frac{{v_F k_F^{} }}{{\varepsilon _F }}} \right)^2
\varepsilon _F \tau $. Finally, Fermi surface can be determined by ARPES,
and all parameters in LLG equation Eq.(\ref{eq:LLG2}) can be obtained.


In summary, we have investigated theoretically the dynamics of magnetization on the surface of a three dimensional topological insulator. We have derived
the Landau-Lifshitz-Gilbert equation in the presence of charge current,
and analyzed the inverse spin-Galvanic effect and ferromegnetic resonance predicting
anomalous features of these phenomena.

This work is supported by Grant-in-Aid for Scientific Research (Grants No.
17071007, 17071005, 19048008 19048015, and 21244053) from the Ministry of
Education, Culture, Sports, Science and Technology of Japan.

\end{document}